\documentclass[%
reprint,
showpacs,preprintnumbers,
amsmath,amssymb,
aps,
pra,
]{revtex4-1}

\usepackage{graphicx}% Include figure files
\usepackage{dcolumn}% Align table columns on decimal point
\usepackage{bm}% bold math
\usepackage{braket}
\usepackage{amsmath}
\begin{document}

%\preprint{APS/123-QED}

\title{Collisional excitation transfer and quenching in Rb($5P$)-methane mixtures}% Force line breaks with \\

\author{M. Alina Gearba}
\email{mirela.gearba@usafa.edu}
\author{Jeremiah H. Wells}
\author{Philip H. Rich}
\author{Jared M. Wesemann}
\author{Lucy A. Zimmerman}
\author{Brian M. Patterson}
\author{Randall J. Knize}
\author{Jerry F. Sell}
\affiliation{Laser and Optics Research Center, Department of Physics, U.S. Air Force Academy, USAF Academy, Colorado 80840, USA}

\date{\today}% It is always \today, today,

\begin{abstract}

We have examined fine-structure mixing between the rubidium $5^{2}P_{3/2}$ and $5^{2}P_{1/2}$ states along with quenching of these states due to collisions with methane gas. Measurements are carried out using ultrafast laser pulse excitation to populate one of the Rb $5^{2}P$ states, with the fluorescence produced through collisional excitation transfer observed using time-correlated single-photon counting.  Fine-structure mixing rates and quenching rates are determined by the time dependence of this fluorescence.  As Rb($5^{2}P$) collisional excitation transfer is relatively fast in methane gas, measurements were performed at methane pressures of $2.5 - 25$ Torr, resulting in a collisional transfer cross section ($5^{2}P_{3/2} \rightarrow 5^{2}P_{1/2}$) of $(4.23 \pm 0.13) \times 10^{-15}$ cm$^{2}$.  Quenching rates were found to be much slower and were performed over methane pressures of $50 - 4000$ Torr, resulting in a quenching cross section of $(7.52 \pm 0.10) \times 10^{-19}$ cm$^{2}$.  These results represent a significant increase in precision compared to previous work, and also resolve a discrepancy in previous quenching measurements.

\end{abstract}

%\pacs{33.50.-j, 34.20.Cf, 32.50.+d}% PACS, the Physics and Astronomy
\maketitle

\section{Introduction}

Initial observations of energy transfer between the fine-structure states of an alkali atom induced by collisions with a buffer gas occurred over a century ago \cite{wood_resonance_1918}.  Experiments carried out in the 1960s to the 1980s measured excitation transfer (mixing) cross sections between alkali $^{2}P_{1/2,3/2}$ states in collisions with inert buffer gases \cite{krause_collisional_1966,jordan_collision-induced_1966,pitre_sensitized_1967,gallagher_rubidium_1968,krause_advances_1975}.  Mixing ($^{2}P_{3/2} \leftrightarrow$ $^{2}P_{1/2}$) and quenching ($^{2}P_{1/2,3/2} \rightarrow$ $^{2}S_{1/2}$) cross sections for alkali atoms in collisions with various molecular gases were also measured during this period \cite{mcgillis_inelastic_1968,hrycyshyn_inelastic_1970,siara_inelastic_1973,phaneuf_52p1/2_1980,krause_advances_1975}.  These collisional processes are a source of current interest due to their relevance in the operation of a diode pumped alkali laser (DPAL), a new class of optically pumped laser whose active medium is an alkali vapor \cite{krupke_resonance_2003,zhdanov_multiple_2008,bogachev_diode-pumped_2012,krupke_diode_2012,zhdanov_review_2013}.  Recent excitation transfer measurements have focused on understanding fine-structure mixing as a function of fine-structure splitting and adiabaticity \cite{eshel_role_2017}, the influence of three-body collisions at high inert gas pressures \cite{sell_enhancement_2010, gearba_temperature_2012,sell_collisional_2012}, and fine-structure transfer in higher-lying alkali $^{2}P$ and $^{2}D$ states \cite{brown_spin-orbit_2012,davila_spin-orbit_2016}.

We present here our efforts to precisely measure the mixing and quenching cross sections for Rb($5 ^{2}P$) states in the presence of methane (CH$_{4}$) gas.  Previous measurements carried out with Rb-CH$_{4}$ mixtures are given in Table I.  The Rb $5 ^{2}P_{3/2} \rightarrow 5 ^{2}P_{1/2}$ collisional mixing cross sections in methane gas (denoted by $\sigma_{21}$) are in good agreement with the exception of Ref. \cite{bulos_collisional_1972}, which is approximately 30\% larger than the other results.  For the $5 ^{2}P_{3/2} \rightarrow 5 ^{2}S_{1/2}$ quenching cross section (denoted by $\sigma_{20}$), the discrepancy in measurements is more dramatic.  The results of Hrycyshyn and Krause \cite{hrycyshyn_inelastic_1970} and Bulos \cite{bulos_collisional_1972} are in agreement, with a notably large uncertainty in the measurement of Ref. \cite{hrycyshyn_inelastic_1970}.  However, a more recent measurement by Zameroski $\textit{et al.}$ \cite{zameroski_study_2009}, placed an upper bound on the combined $5^{2}P_{3/2,1/2} \rightarrow 5^{2}S_{1/2}$ quenching cross section which is more than two orders of magnitude smaller than the previous results.  Measurements of the $5 ^{2}P_{1/2} \rightarrow 5 ^{2}S_{1/2}$ quenching cross section (denoted by $\sigma_{10}$) suggest a smaller quenching cross section for $\sigma_{10}$ compared to $\sigma_{20}$, but do not provide a definitive value.  The experimental temperatures at which each of the measurements were carried out are also listed in Table I.

\begin{table} [b]
\caption{\label{tab:table1} Previous results of experimental mixing and quenching cross sections in Rb-CH$_{4}$ mixtures.}
\begin{ruledtabular}
\begin{tabular}{ccccc}
$\sigma_{21}$  &  $\sigma_{20}$  &  $\sigma_{10}$  &  Temp.  &  Ref. \\
(10$^{-15}$cm$^{2}$)  &  (10$^{-16}$cm$^{2}$)  &  (10$^{-16}$cm$^{2}$)  &  (K)  &   \\
\hline
$4.2 \pm 0.4$  &  $3 \pm 2$  &  $< 1$   &  340  &  \cite{hrycyshyn_inelastic_1970}  \\
$5.3 \pm 0.3$  &  $3.5 \pm 0.2$  &  0   &  293  &  \cite{bulos_collisional_1972}  \\
$3.8 \pm 0.4$  &  $-$  &  $-$    &  310  &  \cite{phaneuf_52p1/2_1980}  \\
$4.1 \pm 0.5$  &  $-$  &  $-$    &  330  &  \cite{rotondaro_role_1998}  \\
$-$  & $\leq 0.019$*   &         &  313  &  \cite{zameroski_study_2009}  \\   
\end{tabular}
\end{ruledtabular}
\begin{flushleft}
*This value is an upper limit of the combined $\sigma_{20}$ and $\sigma_{10}$ cross sections.
\end{flushleft}
\end{table}

Three different experimental techniques were used to obtain the results listed in Table 1.  The most common method \cite{hrycyshyn_inelastic_1970,phaneuf_52p1/2_1980,rotondaro_role_1998} utilizes continuous excitation to one of the fine-structure doublets, with the ratio of fluorescent intensity between the two fine-structure states measured as a function of buffer gas pressure.  The measurements of Bulos \cite{bulos_collisional_1972} incorporate optical pumping techniques to measure the transmitted light intensity in addition to the fluorescent intensity from the fine-structure states.  Both of these experimental methods use relatively low buffer gas pressures not exceeding 20 Torr.  The recent work by Zameroski $\textit{et al.}$ \cite{zameroski_study_2009} measured the time-resolved fluorescence from collisional excitation transfer at temperatures from $ 40 - 130$ $^{\circ}$C and pressures from $50 - 700$ Torr, with significant effects from radiation trapping observed.  The experiments presented here use time-correlated single-photon counting to observe the time dependence of the fluorescence due to collisional excitation transfer after excitation from an ultrafast laser pulse.  The apparatus has been designed to minimize the effects from radiation trapping, while accommodating buffer gas pressures from $0 - 4000$ Torr.

For applications which rely on a buffer gas to transfer population between atomic levels (such as DPAL systems \cite{pitz_recent_2017}, spin-exchange optical pumping \cite{walker_spin-exchange_1997}, and dense gas laser cooling \cite{vogl_laser_2009}), mixing and quenching cross sections are key parameters.  A Rb DPAL typically operates using a high power but low beam quality diode laser to excite the Rb $5^{2}S_{1/2} \rightarrow 5^{2}P_{3/2}$ transition \cite{krupke_diode_2012, zhdanov_review_2013, pitz_recent_2017}.  Buffer gases are included in the alkali vapor to both broaden the atomic transition to more closely match the pump laser bandwidth, and to enable collisional excitation transfer between the $5^{2}P_{3/2}$ and $5^{2}P_{1/2}$ states.  These processes generate a population inversion in the $5^{2}P_{1/2}$ state, with lasing occurring along the $5^{2}P_{1/2} \rightarrow 5^{2}S_{1/2}$ transition with a highly coherent output beam.  Methane is often used as a buffer gas in Rb and Cs DPALs due to its fast fine-structure collisional mixing rates compared to inert gases, and it also does not quench as readily as some other molecular gases \cite{krause_advances_1975}.  As there is significant interest in efficiently scaling DPALs to high powers \cite{bogachev_diode-pumped_2012, pitz_advancements_2016}, various models have been developed to determine the output lasing power under different experimental conditions \cite{hager_three-level_2013, oliker_simulation_2014, cai_analysis_2016, yacoby_analysis_2018, xia_influences_2018}; however, these models rely on having accurate knowledge of mixing and quenching cross sections.

Theoretical calculations of fine-structure collisional transfer cross sections have most often been carried out for alkali-inert gas atom pairs \cite{callaway_inelastic_1965,hidalgo_theory_1972,nikitin_advances_1975}.  For Rb and Cs fine-structure changing collisions in molecular gases, much larger mixing cross sections are observed compared to inert gases which is attributed to energy transfer from electronic to rotational or vibrational states \cite{phaneuf_52p1/2_1980,walentynowicz_inelastic_1974,baylis_rotational_1973}.  Mixing rate measurements across many different molecular gas species suggest a correlation in the probability for collisional transfer as the energy gap between rotational or vibrational states nears that of the alkali fine-structure splitting \cite{pitz_transfer_2011,rotondaro_role_1998}.  Theoretical calculations of the fine-structure collisional transfer cross section have been carried out for Rb-CH$_{4}$ \cite{lawley_collisional_1978} and Cs-CH$_{4}$ \cite{baylis_rotational_1973} using a classical electrostatic model with reasonable agreement to experimental results.  In the case of quenching collisions, molecular gas species are generally assumed to quench more readily than inert gases due to their additional internal energy states.  \textit{Ab initio} potential energy surfaces have recently been calculated for Rb-CH$_{4}$ \cite{heaven_potential_2012}; however, these have not yet to our knowledge been used to determine the fine-structure collisional mixing or quenching cross sections.

\section{Theoretical Background}

\begin{figure}
\includegraphics[width=0.5\textwidth]{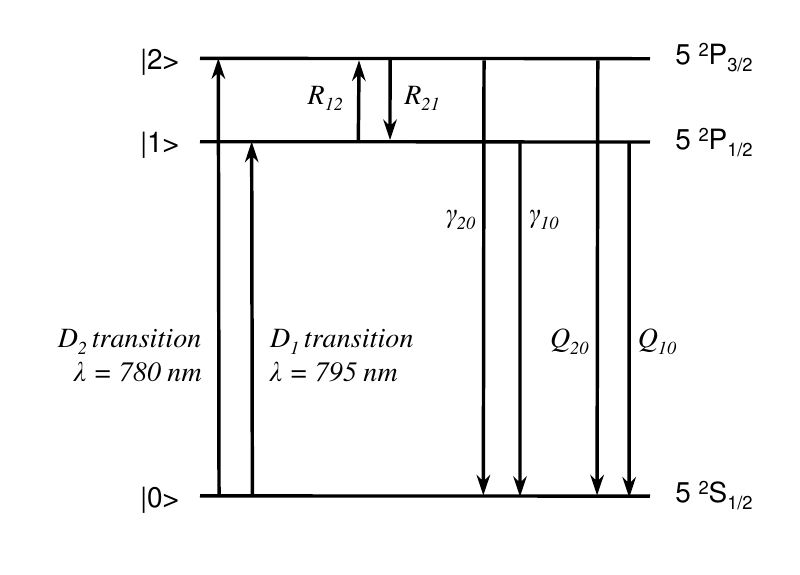}
\caption{\label{fig1-energylevels} Rubidium energy levels involved in this experiment.  Excitation is performed on either the $D_{2}$ transition at 780 nm or the $D_{1}$ transition at 795 nm.  Collisional excitation transfer between Rb $5P$ states is denoted by $R$, the natural radiative decay by $\gamma$, and quenching by $Q$.}
\end{figure}

A schematic of the relevant Rb energy levels and transitions is shown in Figure~\ref{fig1-energylevels}. The $5^{2}S_{1/2}$ ground state is labeled as $\ket{0}$, while the $5^{2}P_{1/2}$ and $5^{2}P_{3/2}$ fine-structure states are labeled as $\ket{1}$ and $\ket{2}$, respectively. Laser excitation is performed from the ground state to either excited state by a pulsed laser and we assume that the excited states are unpopulated prior to a laser pulse. $\gamma_{20}$ and $\gamma_{10}$ represent the radiative decay rates from states $\ket{2}$ and $\ket{1}$, respectively, $R_{21}$ is the collisional excitation transfer (also known as mixing) rate from state $\ket{2}$ to $\ket{1}$, $R_{12}$ is the mixing rate from state $\ket{1}$ to $\ket{2}$, and lastly, $Q_{20}$ and $Q_{10}$ are the collisional quenching (non-radiative) rates from states $\ket{2}$ and $\ket{1}$, respectively, to the ground state $\ket{0}$.

The time evolution of the populations of the two excited states after termination of the laser pulse is described by the following pair of coupled differential equations:
\begin{equation}
\label{eq:rate_eq_2}
\frac{dn_2}{dt}=-(\gamma_{20}+R_{21}+Q_{20})n_2+R_{12}n_1
\end{equation}
\begin{equation}
\label{eq:rate_eq_1}
\frac{dn_1}{dt}=-(\gamma_{10}+R_{12}+Q_{10})n_1+R_{21}n_2,
\end{equation}
where $n_2$ and $n_1$ represent the populations of states $\ket{2}$ and $\ket{1}$, respectively.

If, for example, excitation is performed on the $D_2$ line at 780 nm, we can solve Eq.~(\ref{eq:rate_eq_1}) for $n_2$ and substitute the resulting expression into Eq.~(\ref{eq:rate_eq_2}) to obtain a second-order differential equation in terms of $n_1$: 
\begin{equation}
\label{eq:second_rate_eq_1}
\frac{d^2n_1}{dt^2}+(\alpha_1+\alpha_2)\frac{dn_1}{dt}+(\alpha_1\alpha_2-R_{12}R_{21})n_1=0,
\end{equation}
where $\alpha_1=\gamma_{10}+R_{12}+Q_{10}$ and $\alpha_2=\gamma_{20}+R_{21}+Q_{20}$. The solution to Eq.~(\ref{eq:second_rate_eq_1}) describes the temporal evolution of the $5^{2}P_{1/2}$ population and has the following form:
\begin{equation}
\label{eq:temp_eq_1}
n_1\left(t\right)=Ae^{-s_-t}+Be^{-s_+t},
\end{equation}
where $s_+$ and $s_-$ are given by:
\begin{equation}
\label{eq:coeff_sp_sm}
s_{\pm}=\frac{1}{2}\left[(\alpha_1+\alpha_2){\pm}\sqrt{(\alpha_1-\alpha_2)^2+4R_{12}R_{21}}\right].
\end{equation}
It can be easily observed that $s_+$ and $s_-$ are both positive; therefore, each term of the solution represents an exponential decay. Furthermore, it can also be observed that $s_+>s_-$, resulting in the second term in Eq.~(\ref{eq:temp_eq_1}) representing a faster exponential decay than the first.
Coefficients $A$ and $B$ are determined from the initial conditions, $n_1(0)=0$ and $n_2(0)>0$, for our example of excitation at 780 nm:
\begin{equation}
\label{eq:coeff_A_B}
A=-B=\frac{n_2(0)R_{21}}{\sqrt{(\alpha_1-\alpha_2)^2+4R_{12}R_{21}}}.
\end{equation}

The time evolution of the state $\ket{1}$ population is, therefore, described  by a double-exponential, with the rising portion given by $s_+$, and the decay portion given by $s_-$. If excitation is performed on the $D_1$ line at 795 nm, the set of coupled differential equations with the appropriate initial conditions result in an equation describing the time evolution of state $\ket{2}$ population.

The mixing rates $R_{21}$ and $R_{12}$ are related by the principle of detailed balance:
\begin{equation}
\label{eq:det_bal}
\frac{R_{12}}{R_{21}}=\frac{g_2}{g_1}e^{-\Delta E/k_BT},
\end{equation}
where $g_2=4$ and $g_1=2$ are the degeneracies of states $\ket{2}$ and $\ket{1}$ respectively, $\Delta{E}$ is the difference in energy between these states, namely the fine-structure splitting, $k_B$ is the Boltzmann constant, and $T$ is the temperature measured in Kelvin.

The radiative decay rates $\gamma_{20}$ and $\gamma_{10}$, the inverse of the respective excited state lifetimes, are well known due to several precision measurements of Rb $5^{2}P_{3/2}$ and $5^{2}P_{1/2}$ lifetimes reported in the literature \cite{volz_precision_1996,simsarian_lifetime_1998,gutterres_determination_2002,schultz_measurement_2008}.

Quenching manifests itself as a faster decay to the ground state compared to the radiative decay. We observed no signs of collisional quenching at low pressures, but quenching was clearly present at high pressures. This observation allowed us to perform the study in two different pressure regimes \cite{huennekens_radiation_1987}.

In the low-pressure regime ($<25$ Torr), the quenching rates can be neglected in Eq.~(\ref{eq:temp_eq_1}), which describes the time evolution of the $5^{2}P_{1/2}$ population. Therefore, a fit to the experimental data using Eq.~(\ref{eq:temp_eq_1}) yields the collisional mixing (excitation transfer) rate $R_{21}$ (or $R_{12}$). 

Our study is performed in a vapor cell at a constant temperature, where the atomic velocities follow a Maxwell-Boltzmann distribution. The mixing rate $R_{21}$ and the (velocity averaged) cross-section $\sigma_{21}$ are then related by
\begin{equation}
\label{eq:R_Sigma}
R_{21}=n\sigma_{21}v_{rel},
\end{equation}
where $n$ is the methane density and $v_{rel}$ is the mean relative velocity of the colliding partners given by
\begin{equation}
\label{eq:v_rel}
v_{rel}=\sqrt{\frac{8k_BT}{\pi \mu}},
\end{equation}
with $\mu$ the reduced mass
\begin{equation}
\label{eq:red_mass}
\mu=\frac{m_{Rb}m_{CH_4}}{m_{Rb}+m_{CH_4}}.
\end{equation}

At high pressures ($>50$ Torr), the mixing rates are much higher than the decay rates ($R>>\gamma+Q$). It can be shown that in this regime, the faster rate $s_+$ reduces to \cite{huennekens_radiation_1987}
\begin{equation}
\label{eq:coeff_sp_high}
s_+=R_{12}+R_{21}.
\end{equation}
The time $1/s_+$ required to mix the fine-structure states is, therefore, much shorter than their natural lifetimes of about 27 ns. For example, the experimental value reported in Ref. \cite{rotondaro_role_1998} for the Rb $5^{2}P$ mixing cross-section in methane gives mixing times of 1.1, 0.11, and 0.01 ns at methane pressures of 50, 500 and 4000 Torr, respectively.

As a consequence, in this complete-mixing regime, the three-level system behaves as a quasi-two-level system in which the population ratio of the fine-structure states is fixed by the vapor cell temperature according to the Boltzmann distribution 
\begin{equation}
\label{eq:popratio}
\frac{n_2}{n_1}=\frac{g_2}{g_1}e^{-\Delta{E}/k_BT}. 
\end{equation}
As a result, both fine-structure states decay as a single exponential with the slower rate $s_-$ given by \cite{huennekens_radiation_1987, huennekens_radiation_1983, colbert_radiation_1990}
\begin{equation}
\label{eq:coeff_sm_high}
\begin{split}
s_- &=\frac{1}{2}\Big[(\gamma_{10}+Q_{10}+\gamma_{20}+Q_{20}) \\
& - \frac{(\gamma_{10}+Q_{10}-\gamma_{20}-Q_{20})(R_{12}-R_{21})}{R_{12}+R_{21}}\Big].
\end{split}
\end{equation}

Using Eq.~(\ref{eq:det_bal}), Eq.~(\ref{eq:coeff_sm_high}) then becomes
\begin{equation}
\label{eq:coeff_sm_eff}
s_-=f(\gamma_{10}+Q_{10})+(1-f)(\gamma_{20}+Q_{20}),
\end{equation}
where
\begin{equation}
\label{eq:fraction}
f=\frac{1}{1+\frac{g_2}{g_1}e^{-\Delta{E}/k_BT}}
\end{equation}
is the fraction of population in state $\ket{1}$, while $1-f$ is the fraction of population in state $\ket{2}$. Eq.~(\ref{eq:coeff_sm_eff}) can be re-written in terms of average values for $\gamma$ and $Q$ as
\begin{equation}
\label{eq:coeff_sm_eff_short}
s_-=\gamma_{av}+Q_{av},
\end{equation}
where $\gamma_{av}=f\gamma_{10}+(1-f)\gamma_{20}$ and $Q_{av}=fQ_{10}+(1-f)Q_{20}$.

Under these conditions, a measurement of $s_-$ can only determine the statistically weighted average quenching rate $Q_{av}$, and not the individual quenching rates $Q_{10}$ and $Q_{20}$. The quenching rate is related to the quenching cross section by an expression similar to Eq.~(\ref{eq:R_Sigma}). The slope of the average quenching rate $Q_{av}$ plotted as a function of the methane density $n$ yields the  average quenching cross section $\sigma_Q$, while the $y$ intercept gives the statistically weighted average radiative rate $\gamma_{av}$.

In the complete-mixing regime, a measurement of $s_+$ can lead to another determination of the mixing rate, in principle, but for these high pressures, the rise in the fluorescence with time is much faster than the time response of our detection system and therefore, such a measurement is not possible.

\section{Experimental Setup}

To measure the time dependence of the photons generated due to collisional excitation transfer, we employ the method of time-correlated single-photon counting \cite{oconnor_time-correlated_1984}.  Recent measurements have also utilized an analog signal from the collected fluorescence to measure mixing and quenching rates \cite{zameroski_study_2009, brown_spin-orbit_2012, davila_spin-orbit_2016}; however, we choose to use single-photon counting due to the technique's ability to achieve a high temporal resolution with low noise.  Additional factors in the design of the experimental apparatus include minimizing the effects of radiation trapping, along with the ability to perform measurements over a wide range of buffer gas pressures.

\begin{figure}
\includegraphics[width=0.5\textwidth]{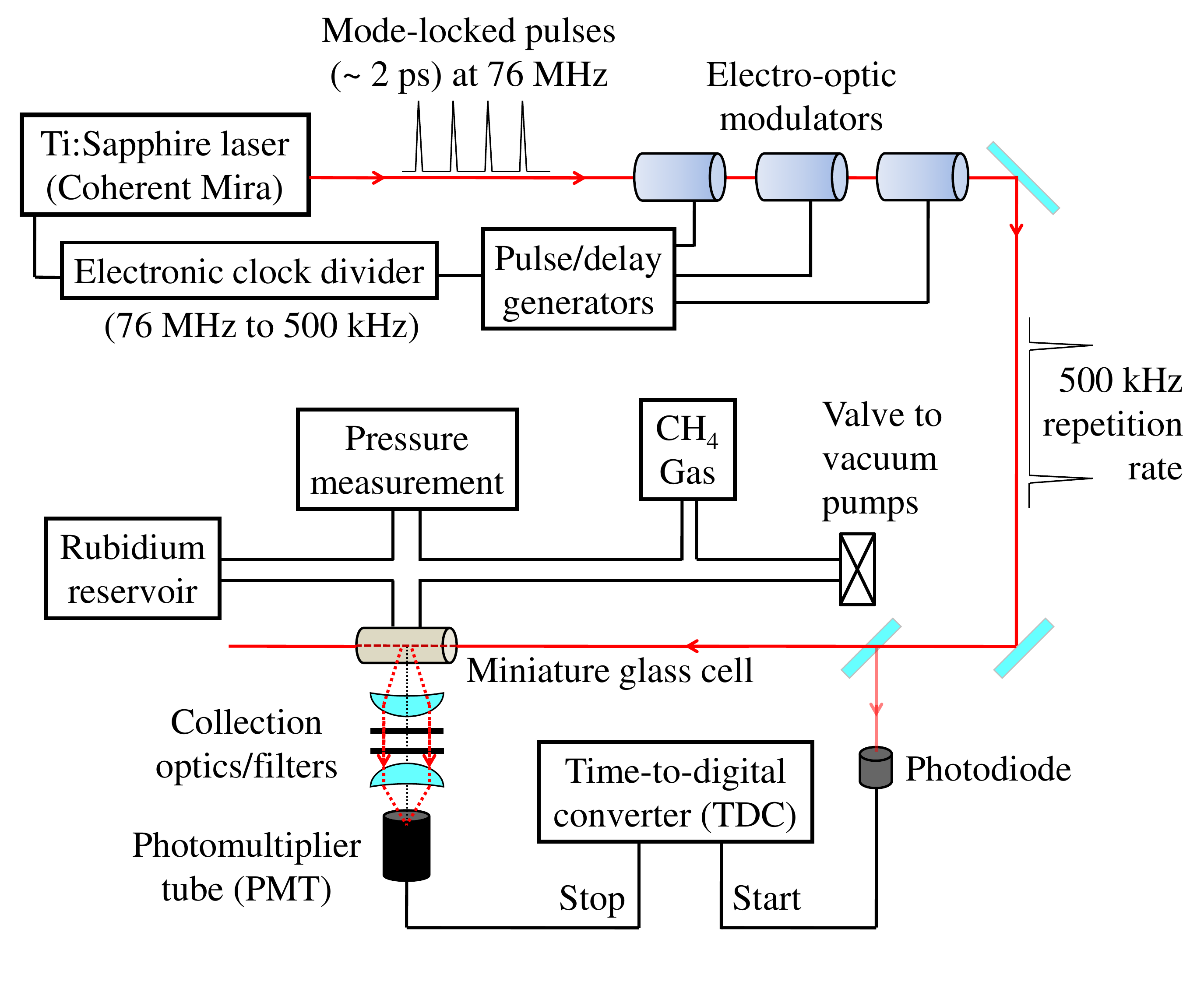}
\caption{\label{fig2-setup}  Schematic diagram of the experimental setup.  Ultrafast laser pulses excite Rb atoms to the state of interest.  Time-correlated single-photon counting is employed to observe the photons emitted due to collisional excitation transfer in time.}
\end{figure}

A schematic diagram of our experimental setup is shown in Fig.~\ref{fig2-setup}.  A mode-locked Ti:S laser (Coherent Mira) is tuned to excite either the Rb $5^{2}P_{3/2}$ state at 780 nm or the $5^{2}P_{1/2}$ state at 795 nm.  This laser has an average output power of 1 W, a pulse repetition rate of 76 MHz, and a pulse duration of approximately 2 ps.  The linewidth of the laser is broad enough to excite all of the hyperfine components of the chosen Rb($5^{2}P$) state, but narrow enough that only one of the fine-structure states is excited.  Three electro-optic modulators (EOMs) are used to lower the laser pulse repetition rate from 76 MHz to 500 kHz in order to provide a 2 $\mu$s window between laser pulses for the observation of atomic fluorescence.  The signal to control the EOMs is generated by sampling the mode-locked optical pulses with a fast photodiode.  These electrical pulses are sent to an electronic frequency divider (Conoptics 305 synchronous countdown) set to divide the input frequency by a factor of 152.  The frequency divider outputs electronic trigger pulses at 500 kHz synchronized to the optical pulses.  These trigger pulses are sent to two pulse/delay generators (Stanford Research Systems DG535 and Berkeley Nucleonics 575) which produce pulses with the appropiate time delay and duration to allow the EOMs (Conoptics 360-80) to pass only a single optical pulse for every 2 $\mu$s  cycle.  The three EOMs in series achieve an extinction ratio between selected and residual optical pulses of greater than $2 \times 10^{4}$:1.

The interaction between the laser pulses and the Rb-CH$_{4}$ gas mixture takes place within a small glass cell attached to a vacuum system.  The vacuum system allows the apparatus to be pumped down to high vacuum levels ($\leq 1 \times 10^{-6}$ Torr) before methane gas is introduced.  Rubidium is contained within a side-arm of the vacuum system and consists of a 1 gram sealed Rb ampoule which is broken under vacuum.  The Rb is heated when it is first introduced into the vacuum system to allow it to migrate throughout the chamber; however, the measurements discussed here all occur at room temperature (298 K).  Methane gas of 99.999$\%$ purity is introduced into the apparatus through a gas handling system to allow precise control of the gas pressure.  The gas pressure is monitored by two capacitance manometers, one for pressures between 1 to 100 Torr (MKS Baratron 626A12TBE) and another for pressures between 50 to 4,000 Torr (MKS Baratron 625D14THAEB).  The glass cell interaction region (Allen Scientific Glass) has a cylindrical geometry with a length of 25 mm and an inner diameter of 2 mm with optical windows attached at each end of the cylinder.  The small inner diameter minimizes both the effects of radiation trapping and the forces present when backfilling the apparatus to high gas pressures.

Atomic fluorescence from collisional excitation transfer is observed orthogonally to the direction of the laser beam propagation.  A 1:1 imaging system ($f/3$) collects the fluorescence and focuses it onto a photon detector.  Two different detectors are used over the course of our experiments; a photomultiplier tube (Hamamatsu R636-10) and a silicon photon avalanche diode (Micro Photon Devices PD-200-CTX).  Multiple filters are inserted into the detection system to allow observation only at 780 nm or 795 nm (with a bandwidth of approximately 10 nm).  The signal from the photon detector is amplified and discriminated and sent to a time-to-digital converter (Agilent Acqiris TC890).  Samples of the 500 kHz laser pulses are detected with a fast photodiode and sent to the time-to-digital converter (TDC).  For every detected event, the TDC precisely measures the time between the incoming laser pulse and the observation of a fluorescence photon.

\section{Results and Discussion}

As discussed in Section II, these experiments are performed in two different pressure regimes.  Measurements of the Rb $5^{2}P_{3/2} \leftrightarrow 5^{2}P_{1/2}$ collisional excitation transfer (mixing) rates are performed in the low-pressure regime ($2.5-25$ Torr), while measurements of the $5^{2}P_{3/2,1/2} \rightarrow 5^{2}S_{1/2}$ quenching rates are performed in the high-pressure regime ($50-4000$ Torr).

\subsection{Rb $5^{2}P$ fine-structure collisional transfer cross sections in methane gas}

\begin{figure}[b]
\includegraphics[width=0.5\textwidth]{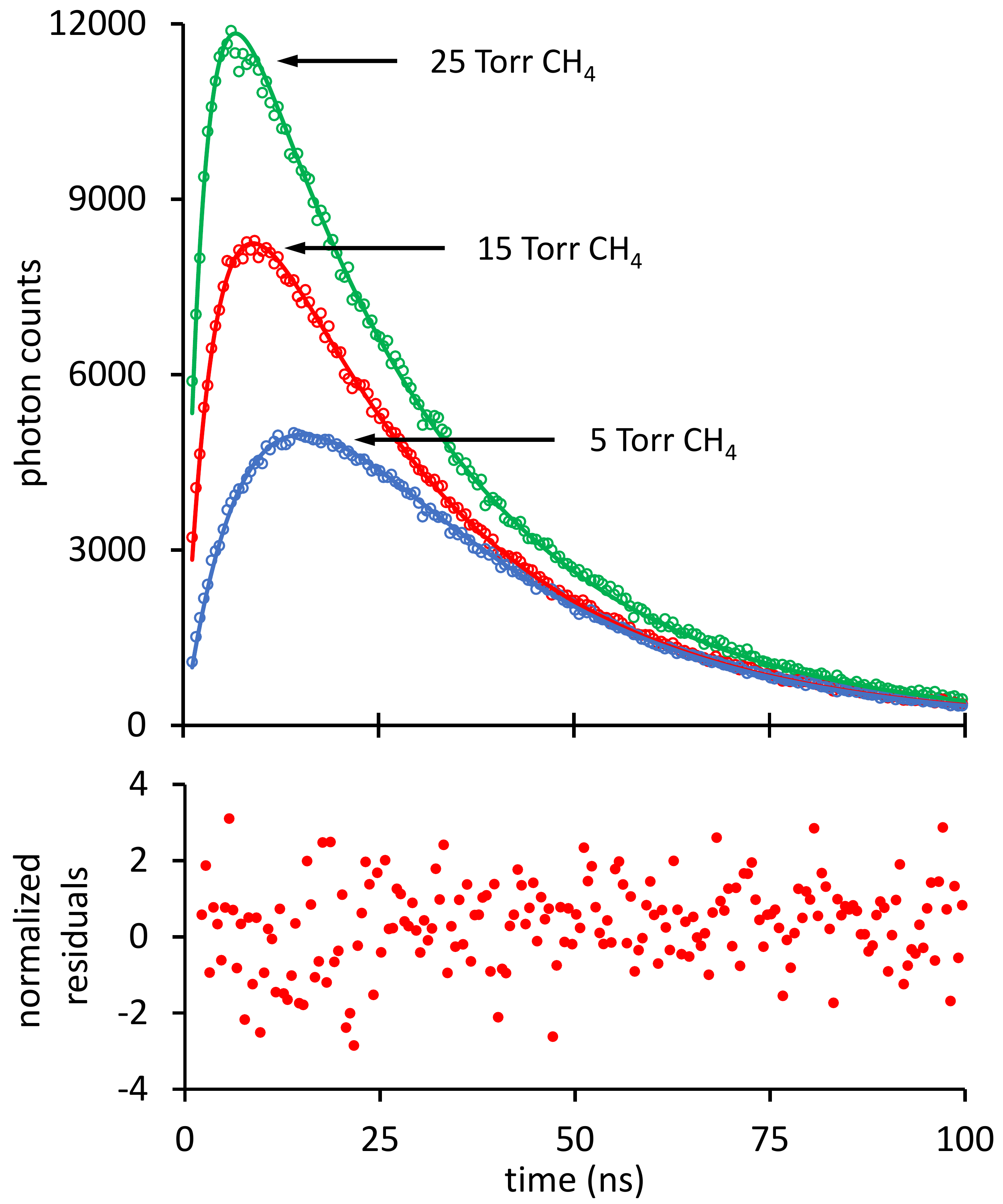}
\caption{\label{fig3-mixingfluorescence}  The distribution of photons in time observed at 780 nm (open circles) after excitation of Rb at 795 nm with methane gas present at the pressures shown.  The time axis is calibrated with respect to the laser pulse, and the solid lines are fits to the data with a functional form given by Eq.(4).  The lower plot illustrates the normalized residuals obtained from fitting the 15 Torr CH$_{4}$ data set. }
\end{figure}

Figure~\ref{fig3-mixingfluorescence} shows a typical fluorescence histogram recorded in the low-pressure regime using laser excitation at 795 nm and detecting the fluorescence due to collisonal excitation transfer at 780 nm.  Fluorescence histogram are typically recorded for $10 - 20$ minutes, using 0.5 ns wide time-bins.  The time axis is calibrated based on the observation of the laser excitation pulse in the absence of methane buffer gas (and with the filters for scattered laser light removed). Each data set is fit to Eq.~(\ref{eq:temp_eq_1}), with the coefficients given by Eqs.~(\ref{eq:coeff_sp_sm}) and (\ref{eq:coeff_A_B}). The quenching rates $Q_{10}$ and $Q_{20}$ are neglected in this fit and the mixing rates $R_{12}$ and $R_{21}$ are related by Eq.~(\ref{eq:det_bal}). Under these conditions, the time evolution of the $5^{2}P_{3/2}$ population at a given pressure is described by three independent parameters: the collisional mixing rate $R_{21}$, a dimensionless scaling factor, and a constant background. Data fitting is performed in MATLAB using a Levenberg-Marquardt nonlinear least-squares fitting routine. Also shown in Fig.~\ref{fig3-mixingfluorescence} are the normalized residuals as a result of a typical data set fit.

\begin{figure}[b]
\includegraphics[width=0.5\textwidth]{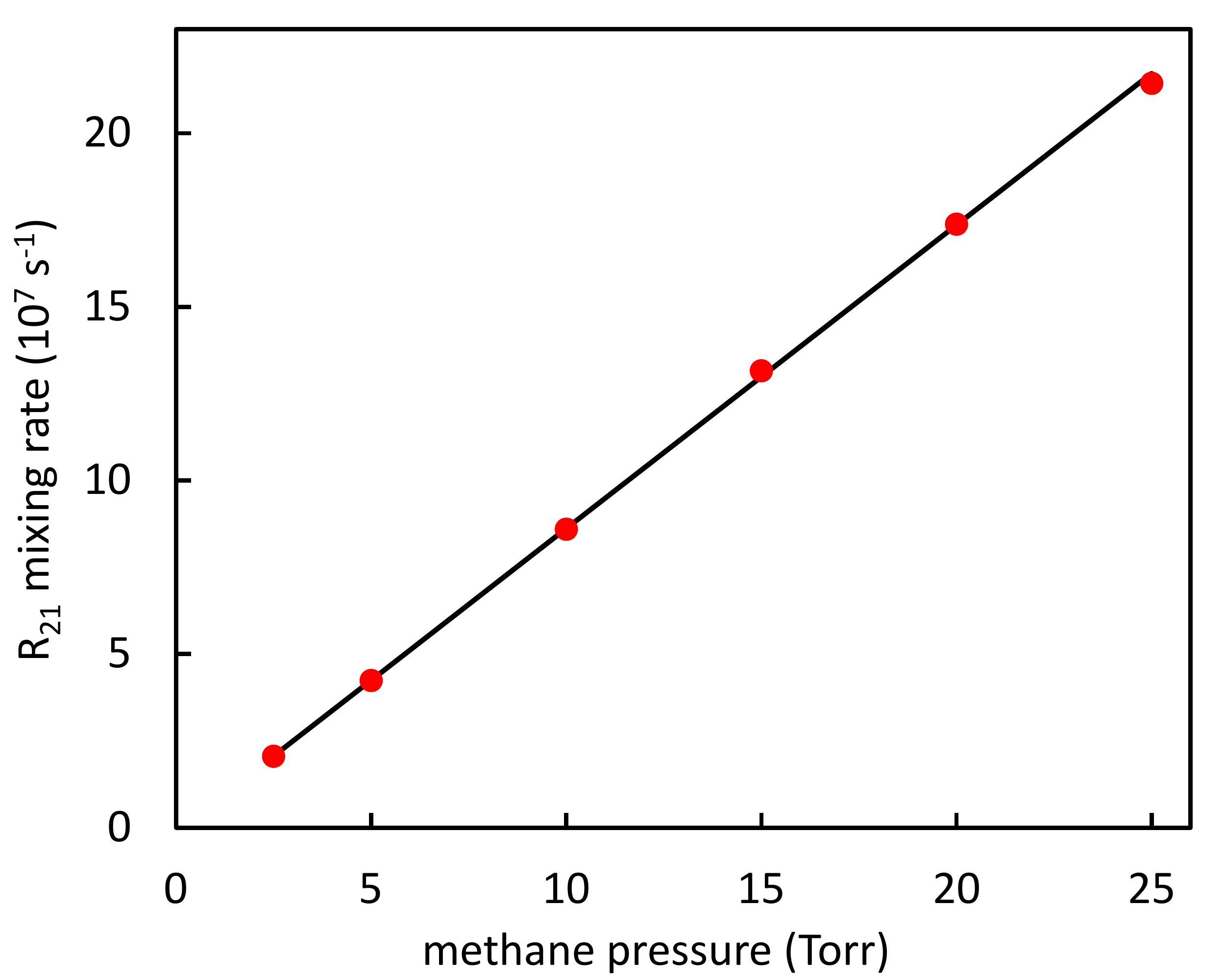}
\caption{\label{fig4-mixingrates}  Pressure dependence of the Rb($5^{2}P$) mixing rate, $R_{21}$, over a range of methane pressures from 2.5 to 25 Torr.  This data set was taken with laser excitation at 795 nm and fluorescence detection at 780 nm.  A linear fit to the data points (solid line) results in a value of $\sigma_{21} = (3.95 \pm 0.02) \times 10^{-15}$ cm$^{2}$ (statistical error only).  Error bars are not shown as they are within the size of the data points.}
\end{figure}

To determine the fine-structure collisional transfer cross section, fluorescence curves are obtained at multiple methane pressures between 2.5 to 25 Torr, as illustrated in Fig.~\ref{fig4-mixingrates}.  For this particular data set, laser excitation is performed at 795 nm and fluorescence detection at 780 nm.  To check for any inconsistencies in our measurements, we also acquire data under the reverse conditions of laser excitation at 780 nm and fluorescence detection at 795 nm.  In either case, the principle of detailed balance, Eq.(\ref{eq:det_bal}), fixes the relationship between $R_{21}$ and $R_{12}$.  For the data set shown in Fig.~\ref{fig4-mixingrates}, a linear fit to the data is carried out and used with Eq.~(\ref{eq:R_Sigma}), resulting in a mixing cross section of $\sigma_{21} = (3.95 \pm 0.02) \times 10^{-15}$ cm$^{2}$, with the given error representing the statistical error only.  Multiple data sets similar to those in Fig.~\ref{fig4-mixingrates} are acquired for both excitation/detection wavelength combinations.

In addition to statistical errors, various systematic effects are present which contribute to the uncertainty of our measurements.  A listing of these errors and their contributions are given in Table II.  Several of these systematic effects have been described in detail previously \cite{sell_collisional_2012}; here we briefly discuss how we determined the various sources of error.

\begin{table}
\caption{\label{tab:table2} Summary of error contributions in the determination of the fine-structure collisional transfer cross section, $\sigma_{21}$.}
\begin{ruledtabular}
\begin{tabular}{l c}
Source of uncertainty  &   Error ($\%$)   \\
\hline
Pressure measurement 				    						&    $\pm$ 	 0.7	\\
Time calibration of TDC  										&    $\leq$  0.01	\\
Height uniformity of TDC histogram		   			 			&    $\leq$  0.3	\\
Time calibration to laser pulse (weighted ave.)					&    $\pm$	 2.0	\\
\hspace{50 pt}	PMT detector $\rightarrow$ $\pm$ $5.7$ $\%$   	&                   \\
\hspace{50 pt}	SPAD detector $\rightarrow$ $\pm$ $2.1$ $\%$  	& 					\\
EOM pulse selection  											&    $\leq$  0.2	\\
Pulse pileup   													&    $\leq$  0.4	\\
Truncation error: beginning of data fit   						&    $\pm$ 	 1.8	\\
\hspace{72 pt}   end of data fit								&    $\leq$  0.4    \\
Radiation trapping (correction of $+ 2.1$ $\%$) 				&    $\pm$ 	 1.0	\\
Statistical error    				    						&    $\pm$   0.23 	\\
\hline
Total error	(combined in quadrature)							&    $\pm$	 3.0	\\
\end{tabular}
\end{ruledtabular}
\end{table}

Small systematic errors include those arising from pressure measurements, the time-to-digital converter electronics, pulse pileup effects, and the operation of the electro-optic modulators.  In the case of methane pressure measurements, errors can occur both from the instrument uncertainty of the capacitance manometers, along with drifts in the pressure during data collection.  The TDC electronics also have two sources of error, the first coming from how accurate the TDC is calibrated in time, and the second from how uniform in height are the various time-bins of the TDC histogram.  To check the TDC time calibration, we purposefully allow the detection system to observe the full 76 MHz optical pulse train and compare the time between detected pulses to that obtained from a precise measurement of the repetition rate of the laser.  The height of the various time-bins of the histogram from the TDC are analyzed using a random (in time) light source which should equally populate all time-bins.  We verified that any nonlinearities in the height of the time-bins results in a measurement uncertainty of $\leq 0.3\%$.  Pulse pileup refers to the effect that after a photon is detected, a dead-time occurs within the detector and electronics, which will not detect a subsequent photon until after this dead-time has passed \cite{oconnor_time-correlated_1984}.  To keep the uncertainty from this effect to a level of $\leq 0.4 \%$, the photon counting rate is kept below $2 \times 10^{4}$ s$^{-1}$ during mixing rate measurements.  The electro-optic modulators used in this experiment also cause a small systematic effect as they do not completely extinguish unwanted laser pulses.  To reduce the uncertainty from this effect, three EOMs are used in series to achieve extinction ratios of $\geq 2 \times 10^{4}:1$, which results in an uncertainty from the background laser light of $\leq 0.2\%$.

The largest source of uncertainty in our measurements results from calibrating the time axis of the fluorescence histograms to the laser pulse.  The time axis is not initially calibrated as the starting and stopping signals to the TDC come from different detectors which have unique optical path lengths and electronics.  Typically, at the beginning and end of a data run, when no buffer gas is present, scattered laser light is allowed into the detection system to record a histogram of the incoming laser pulse.  These data sets are fit to determine the position of the laser pulse in time, and this value is used to calibrate the origin of the time axis of the fluorescence histograms.  To improve the uncertainty of this measurement, two photon detection systems were used during the course of our experiments: a photomultiplier tube with a timing resolution of approximately 1 ns (FWHM), and a silicon photon avalanche diode with a quoted timing resolution of 35 ps (FWHM).  When combined with our detection electronics and TDC, the PMT system achieved fits of the laser pulse in time with an uncertainty of $\pm 250$ ps, while the SPAD achieved an uncertainty of $\pm 100$ ps.  We found that these uncertainties translated into errors in the fine-structure collisional transfer cross section of $\pm 5.7 \%$ and $\pm 2.1 \%$ using the PMT and SPAD detectors, respectively.  Results of data sets from both detection systems are combined using a weighted average in the final determination of the fine-structure collisional transfer cross section.

Two additional systematic effects considered in our analysis are radiation trapping and the truncation of data points as part of the fitting procedure.  Radiation trapping refers to the re-absorption of fluorescence photons before they are able to escape the atomic sample \cite{molisch_radiation_1998}.  This effect delays the observation of fluorescence photons, resulting in a smaller mixing rate than would otherwise be measured.  We minimize this effect by having a very small path length (1 mm) for photons to traverse before exiting the atomic sample, along with performing measurements at only 298 K.  A small correction of $2.1\%$ is applied as a consequence of this effect, and is calculated based on the photon absorption probability in Rb \cite{sell_collisional_2012}, taking into account the pressure broadening due to methane gas \cite{rotondaro_collisional_1997}.  Truncation errors refer to a systematic shift in the measured mixing rate depending on the range of data that is fit.  This effect is much more pronounced at the beginning of the data fit when the fluorescence signal is quickly rising and scattered photons from the laser excitation pulse may also be present.  We typically exclude the first $1-2$ ns of data points to avoid scattered photons from the initial laser pulse, with a reasonable range of starting points for the data fit resulting in the uncertainty shown in Table II.  Our data fits extend to typically 500 ns, and we find systematic shifts of no more than $\pm 0.4\%$ when ending the data fits between 200 ns and 2 $\mu$s.

Taking into account statistical and time calibration uncertainties, we measure values of $\sigma_{21}=(4.22 \pm 0.12) \times 10^{-15}$ cm$^{2}$ and $\sigma_{21}=(4.06 \pm 0.13) \times 10^{-15}$ cm$^{2}$ using laser excitation at 780 nm and 795 nm, respectively.  These results are combined using a weighted average and after including the rest of the systematic errors listed in Table II and the radiation trapping correction we achieve our final result of $\sigma_{21}=(4.23 \pm 0.13) \times 10^{-15}$ cm$^{2}$.

\begin{figure}
\includegraphics[width=0.5\textwidth]{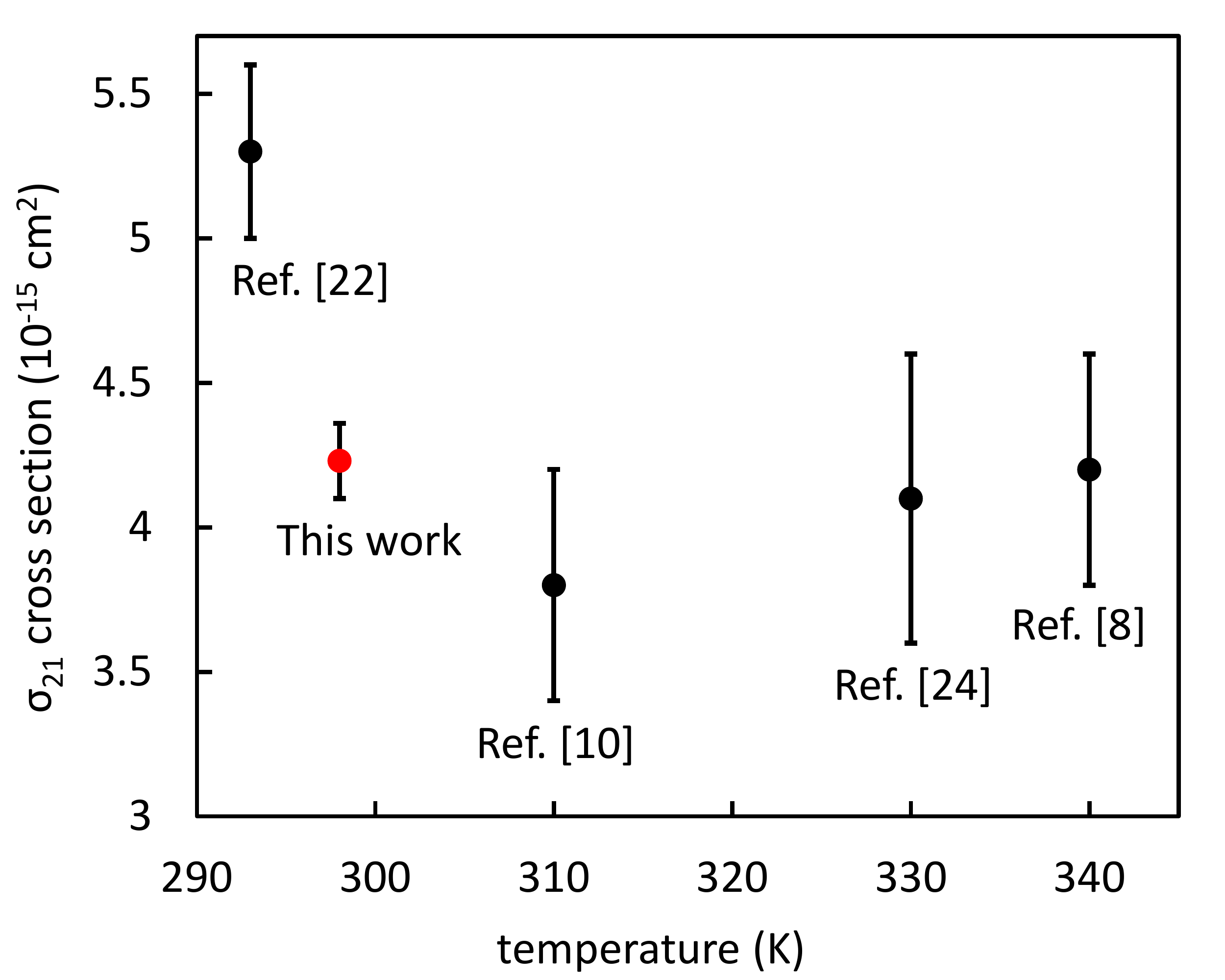}
\caption{\label{fig5-mixingratecompare}  Comparison of the $\sigma_{21}$ cross section determined in this work (red data point at a temperature of 298 K) with the previous determinations given in Table I.}
\end{figure}

Figure~\ref{fig5-mixingratecompare} illustrates our result for the collisional excitation transfer (mixing) cross section for Rb($5^{2}P$) states in methane gas in comparison with the previous measurements listed in Table I.  The results are plotted as a function of the temperature at which the measurements were performed.  We note a slight decrease in the mixing cross section is to be expected as the temperature increases, according to previous temperature dependent measurements \cite{phaneuf_52p1/2_1980}.  Taking this into consideration, we find our results to be in agreement with previous determinations obtained using the sensitized fluorescence technique \cite{hrycyshyn_inelastic_1970,phaneuf_52p1/2_1980,rotondaro_role_1998}, while achieving significantly improved experimental uncertainties.  The studies based on optical pumping techniques in addition to measuring fluorescence give values approximately $25\%$ larger than our results \cite{bulos_collisional_1972}, which cannot be explained by temperature dependence alone.

Further improvements in experimental precision could be obtained using the techniques described in this work primarily through increased photon timing resolution.  Use of the SPAD detector clearly resulted in more precise measurements of the laser pulse arrival time compared to the PMT detector, which is the largest source of error in our measurements.  Increased photon timing resolution would also provide more data points during the relatively quick rise in the fluorescence from collisional excitation transfer, likely decreasing the uncertainty from the truncation of data points at the beginning of the data fit.  While fine-structure collisional transfer cross sections have been measured for many different alkali-buffer gas combinations using the sensitized fluorescence technique, the experiments described here are particularly well suited when high precision measurements are sought.

\subsection{Rb $5^{2}P$ collisional quenching cross sections in methane gas}

\begin{figure}
\includegraphics[width=0.5\textwidth]{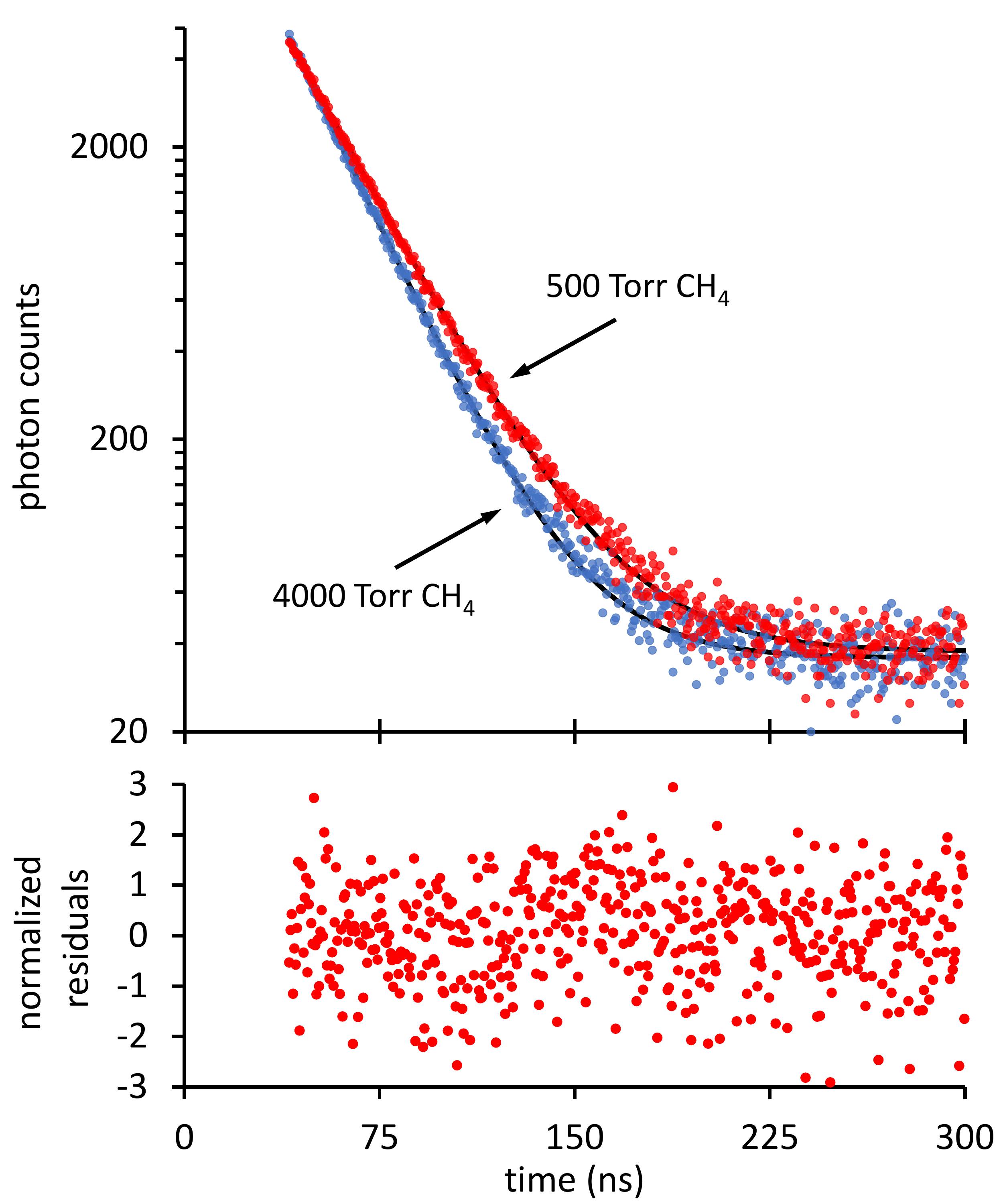}
\caption{\label{fig6-quenchingfluorescence}  The distribution of photons in time observed at 795 nm after excitation of Rb at 780 nm with methane gas present at the pressures shown.  The upper plot is presented on a semilogarithmic scale with fits to the data (an exponential decay with a constant background) shown as solid lines.  The lower plot illustrates the normalized residuals obtained from fitting the 500 Torr CH$_{4}$ data set. }
\end{figure}

Fluorescence data in the high pressure regime of Rb-CH$_{4}$ mixtures is illustrated in Fig.~\ref{fig6-quenchingfluorescence}, with laser excitation carried out at 780 nm and fluorescence detection at 795 nm.  At these high methane pressures, the rise in the collisionally induced fluorescence is nearly instantaneous compared to the time response of the detection system.  As we can no longer measure mixing rates at these high pressures, we focus only on the decay portion of the fluorescence curve.  We note that this also removes the requirement of calibrating the origin of the time axis to the laser pulse.  Data fitting is performed in OriginPro using a Levenberg-Marquardt fitting algorithm to the functional form,
\begin{equation}
n(t)=Ce^{-(s_{-})t}+D,
\end{equation}
where $C$ is a scaling factor which depends on the starting point of the fit, $s_{-}$ is the decay rate, and $D$ is the background counting rate.  For the data presented in Fig.~\ref{fig6-quenchingfluorescence}, $s_{-}$ is measured to be $(3.731 \pm 0.008) \times 10^{7}$ s$^{-1}$ at 500 Torr CH$_{4}$, and $(4.314 \pm 0.011) \times 10^{7}$ s$^{-1}$ at 4000 Torr, with the stated uncertainty representing only the statistical error.  For reference, in the absence of quenching, Eq.~(\ref{eq:coeff_sm_eff}) leads to a statistically weighted average radiative rate of $\gamma_{av} = 3.69 \times 10^{7}$ s$^{-1}$, a deviation of approximately $1\%$ from the 500 Torr result.  The quenching rate $Q_{av}$ is determined by subtracting $\gamma_{av}$ from $s_{-}$, according to Eq.~(\ref{eq:coeff_sm_eff_short}).  The normalized residuals in Fig.~\ref{fig6-quenchingfluorescence} result from fitting the 500 Torr data set.

\begin{figure}
\includegraphics[width=0.5\textwidth]{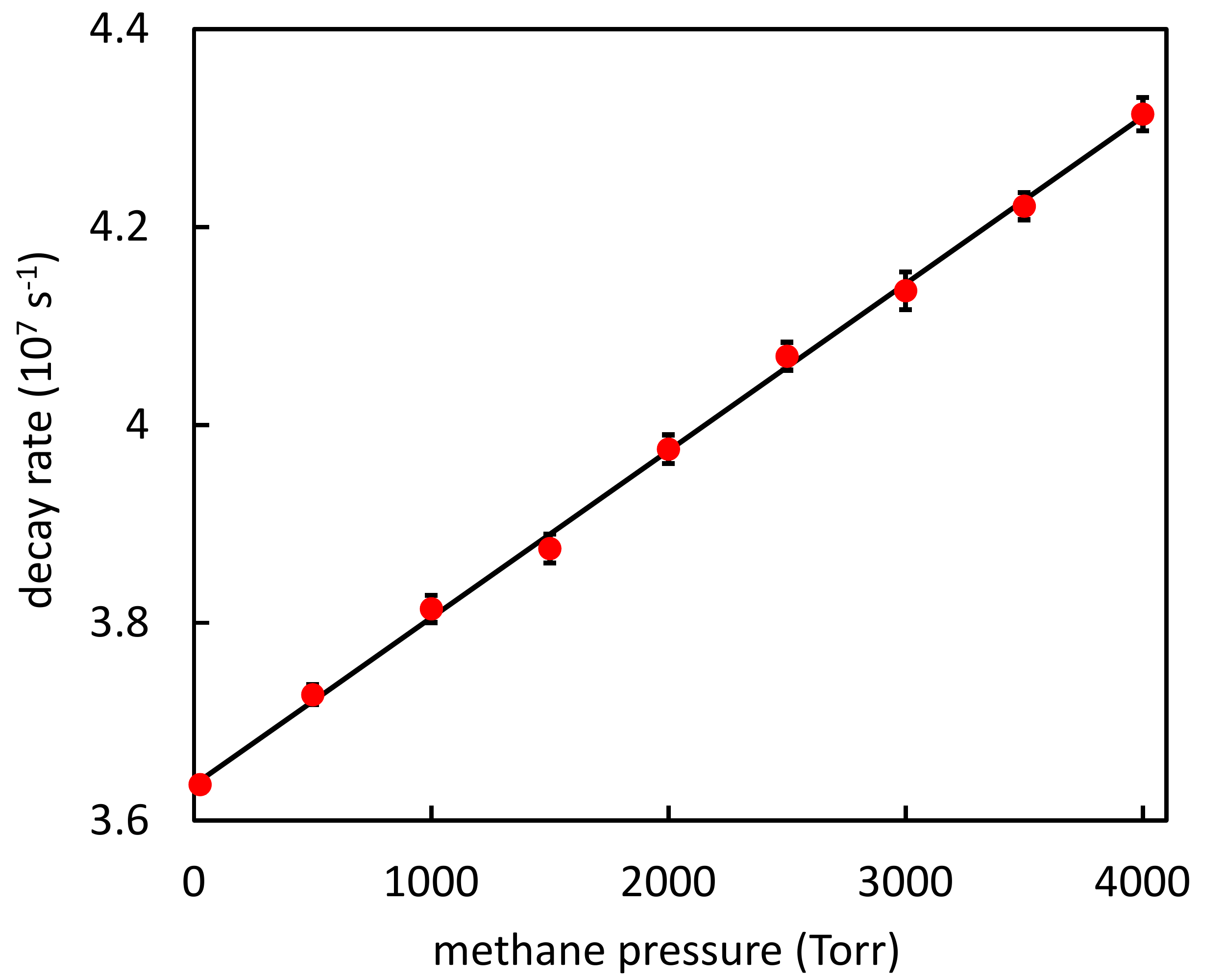}
\caption{\label{fig7-decayrates}  Pressure dependence of the Rb $5P_{3/2}$ decay rate over a range of methane pressures from 50 to 4000 Torr.  This data set was taken with laser excitation at 795 nm and fluorescence detection at 780 nm.  A linear fit to the data points (solid line) results in a value of $\sigma_{Q} = (7.60 \pm 0.11) \times 10^{-19}$ cm$^{2}$ (statistical error only).}
\end{figure}

The collisional quenching cross section is determined by measuring quenching rates over a range of methane gas pressures, as shown in Fig.~\ref{fig7-decayrates}, using laser excitation at 795 nm and fluorescence detection at 780 nm.  A linear fit to this data results in a statistically weighted average quenching cross section of $\sigma_{Q} = (7.60 \pm 0.11) \times 10^{-19}$ cm$^{2}$, which includes statistical error only.  The uncertainties in the quenching cross section from systematic effects are summarized in Table III.  Effects such as the TDC time calibration and height linearity, pressure measurements, and EOM pulse selection are mostly unchanged from the collisional mixing cross section measurement.  The truncation error is much smaller for quenching measurements as the exponential decay is largely unaffected by the starting point of the data fit.  Pulse piluep was substantially reduced in these experiments by keeping the photon counting rate at no more than 2,500 s$^{-1}$.  Lastly, the high methane gas pressures used in these measurements resulted in greatly increased pressure broadening and a corresponding decrease in the probability for radiation trapping to occur.

\begin{table}
\caption{\label{tab:table3} Summary of error contributions in the determination of the quenching cross section, $\sigma_{Q}$.}
\begin{ruledtabular}
\begin{tabular}{l c}
Source of uncertainty  &   Error ($\%$)   \\
\hline
Pressure measurement 				    		&    $\pm$ 	  0.5	\\
Time calibration of TDC  						&    $\leq$   0.01	\\
Height uniformity of TDC histogram		     	&    $\leq$   0.3	\\
EOM pulse selection  							&    $\leq$   0.4	\\
Pulse pileup   									&    $\leq$   0.05  \\
Truncation error: beginning of data fit   		&    $\pm$    0.5   \\
Radiation trapping                           	&    $\leq$   0.2 	\\
Statistical error    				    		&    $\pm$    1.0	\\
\hline
Total error										&    $\pm$	  1.3	\\
\end{tabular}
\end{ruledtabular}
\end{table}

Quenching rate measurements are also carried out with laser excitation at 780 nm and fluorescence detection at 795 nm.  Results for the two cases are in agreement, with $\sigma_{Q} = (7.45 \pm 0.10) \times 10^{-19}$ cm$^{2}$ using $D_{2}$ line excitation, and $\sigma_{Q} = (7.60 \pm 0.11) \times 10^{-19}$ cm$^{2}$ using $D_{1}$ line excitation, including statistical error only.  These results are combined using a weighted mean and after inclusion of the systematic errors listed in Table III give a final value of $\sigma_{Q} = (7.52 \pm 0.10) \times 10^{-19}$ cm$^{2}$.

This measurement is in agreement to the result of Zameroski $\textit{et al.}$ \cite{zameroski_study_2009}, which provided a bound of $\sigma_{Q} \leq 1.9 \times 10^{-18}$ cm$^{2}$, while clearly differing from the other values listed in Table I.  Even considering this rather small quenching cross section, our technique was able to achieve a direct measurement with an experimental uncertainty of $1.3\%$. Experimental differences compared to Ref. \cite{zameroski_study_2009} include the use of time-correlated single-photon counting, the wide range of methane gas pressures used, and the negligible influence radiation trapping had on our results.  An uncertainty of 67$\%$ is quoted for the $\sigma_{20}$ measurement by Hrycyshyn and Krause \cite{hrycyshyn_inelastic_1970}, indicating a lack of sensitivity to the quenching effect in the case of Rb-CH$_{4}$.  The result of Bulos \cite{bulos_collisional_1972} quotes a smaller uncertainty of approximately $6\%$ in the measurement of $\sigma_{20}$, and a quenching cross section of 0 for $\sigma_{10}$.

Based on our measurements, the Rb $5^{2}P$ fine-structure collisional transfer cross section is more than three orders of magnitude larger than the collisional quenching cross section, indicating the fine-structure states are highly mixed before any significant quenching can occur.  Under these conditions, quenching is accurately described by a single parameter, the statistically weighted average quenching cross section.  We note that other molecular gases such as Rb-N$_{2}$ exhibit the inverse condition of larger quenching than mixing cross sections \cite{hrycyshyn_inelastic_1970}, along with many cases where the mixing and quenching cross sections are comparable \cite{krause_advances_1975}.  This large difference between mixing and quenching cross sections in Rb-CH$_{4}$ may be useful in applications such as diode pumped alkali lasers, where fast mixing between fine-structure states is necessary, but quenching from these states should be minimal.

\section{Conclusion}

Using ultrafast laser pulse excitation and time-correlated single-photon counting, we have determined the fine-structure mixing and quenching cross-sections for Rb($5^{2}P$) states due to collisions with methane (CH$_{4}$) gas.  Our measurements of the fine-structure collisional mixing cross section are in agreement with previous results, while achieving significantly improved experimental uncertainties through observation of the temporal evolution of the fluorescence due to collisional excitation transfer.  Further improvements in the precision of mixing rate measurements may be achieved with this experimental technique by increasing the photon timing resolution of the detection system.  

Measurements of the Rb(5$^{2}P$) collisional quenching rates were performed over methane pressures of $50 - 4000$ Torr, resulting in a quenching cross section more than three orders of magnitude smaller than the fine-structure collisional mixing cross section.  These measurements resolve a discrepancy between previously reported quenching results, while also demonstrating the high precision possible with this experimental technique even when measuring small quenching cross sections.  The experimental methods described here can readily be used to measure many other alkali-buffer gas quenching rates, including temperature dependent cross sections.  As few quenching cross sections this small have been measured, additional studies across different molecular buffer gases may give further insight into the energy transfer mechanism.  The findings of this study are also relevant for alkali laser development, as well as for understanding collisional processes in alkali-buffer gas systems.

\begin{acknowledgments}
Support for this work by the National Science Foundation (Grant Nos. 1531107 and 1708165), the Air Force Office of Scientific Research, and the Society of Physics Students is gratefully acknowledged.
\end{acknowledgments}

\end{document}